\documentclass[superscriptaddress, twocolumn, notitlepage, amsmath, amssymb, aps, prl, 10pt, letterpaper]{revtex4-1}

\usepackage{graphicx,times,color,hyperref,tikz}
\usetikzlibrary{backgrounds,external}
\hypersetup{breaklinks}
\begin{document}

\title{Local \texorpdfstring{$\mathcal{PT}$}{PT} symmetry violates the no-signaling principle}

\author{Yi-Chan Lee}
\email{xellosslee@gmail.com}
\affiliation{Physics Department, National Tsing-Hua University, Hsinchu City 300, Taiwan}
\affiliation{Centre for Quantum Computation \& Intelligent Systems, Faculty of Engineering and Information Technology, University of Technology, Sydney, NSW 2007, Australia}

\author{Min-Hsiu Hsieh}
\affiliation{Centre for Quantum Computation \& Intelligent Systems, Faculty of Engineering and Information Technology, University of Technology, Sydney, NSW 2007, Australia}

\author{Steven T. Flammia}
\affiliation{School of Physics, University of Sydney, Sydney, NSW 2006, Australia}

\author{Ray-Kuang Lee}
\affiliation{Physics Department, National Tsing-Hua University, Hsinchu City 300, Taiwan}
\affiliation{Institute of Photonics Technologies, National Tsing-Hua University, Hsinchu City 300, Taiwan}

\date{\today}

\begin{abstract}
Bender \emph{et al.}~\cite{Bender1998} have developed $\mathcal{PT}$-symmetric quantum theory as an extension of quantum theory to non-Hermitian Hamiltonians. We show that when this model has a local $\mathcal{PT}$ symmetry acting on composite systems it violates the non-signaling principle of relativity. Since the case of global $\mathcal{PT}$ symmetry is known to reduce to standard quantum mechanics~\citep{Mostafazadeh2001, *Mostafazadeh2002, *Mostafazadeh2002a, *Mostafazadeh2003, *Mostafazadeh2004, *Mostafazadeh2007}, this shows that the $\mathcal{PT}$-symmetric theory is either a trivial extension or likely false as a fundamental theory. 
\end{abstract}

\maketitle

The Hermiticity of Hamiltonians---and indeed observables in general---is one of the fundamental postulates of quantum mechanics. There are two reasons for this restriction: first, a Hermitian Hamiltonian guarantees that the energy of the physical system described by it is always real. Second, based on the Schr\"{o}dinger equation, the Hermiticity implies that the time-evolution operator generated by a Hamiltonian is unitary, which ensures conservation of probabilities for the time-evolved quantum state. 

Nonetheless, non-Hermitian Hamiltonians are still useful in theoretical work and are a mathematical tool for studying open quantum systems in nuclear physics~\citep{Feshbach1958} or quantum optics~\citep{Plenio1998}, among others. In these fields, the whole physical system is still considered to obey conventional quantum mechanics, and the non-Hermitian Hamiltonian only comes out as an effective subsystem within a projective subspace. 

In 1998, Bender and colleagues proposed a class of non-Hermitian Hamiltonians with a real energy spectrum as a fundamental, non-effective model beyond standard quantum theory~\citep{Bender1998}. By redefining the inner product, the time evolution operator generated by such a Hamiltonian could be unitary \citep{Bender2002}. Their proposal reveals the possibility to remove the restriction of Hamiltonians from Hermiticity to a weaker parity-time ($\mathcal{PT}$) symmetry, where parity-time means spatial reflection and time reversal. In other words, it might be possible to have a physical system described by a non-Hermitian Hamiltonian. They showed that when the eigenstates of a $\mathcal{PT}$-symmetric system are also $\mathcal{PT}$ symmetric, the energy eigenvalues are always real. When the eigenstates are no longer $\mathcal{PT}$ symmetric, the energy becomes complex and is called spontaneous ($\mathcal{PT}$) symmetry breaking. 

This proposal led to a flurry of activity investigating the strange properties of $\mathcal{PT}$-symmetric Hamiltonians. Especially in optical systems, since the paraxial equation is equivalent to Schr\"{o}dinger's equation, various \emph{effective} models were proposed to simulate $\mathcal{PT}$-symmetric Hamiltonian dynamics~\cite{El-Ganainy2007}. A $\mathcal{PT}$-symmetric Hamiltonian was successfully simulated in optics experiments by using coupling optical channels in 2010, and the spontaneous breaking of $\mathcal{PT}$ symmetry was also observed in this system~\cite{Ruter2010}. Besides these discoveries, many optical applications of $\mathcal{PT}$-symmetric Hamiltonians were also proposed, such as unidirectional optical valves~\cite{Ramezani2010}, perfect laser absorbers~\cite{Longhi2010}, unidirectional invisible media~\cite{Lin2011}, and spatial optical switches~\cite{Nazari2011}. The applications of $\mathcal{PT}$-symmetric Hamiltonians in these optical systems are all classical and, to the extent that they were realized, were effective models. However, in the quantum regime Bender and others proposed two interesting applications related to quantum computation: ultrafast quantum state transformation~\citep{Bender2007} and quantum state discrimination with single-shot measurement~\citep{Bender2013}, which also inspired much investigation of ``shortcut'' quantum time evolution~\citep{Ibanez2011, Torosov2013}.

It is well known that in conventional quantum mechanics the time to evolve between two orthogonal states is limited by the uncertainty principle~\citep{Anandan1990, Margolus1998}, and only orthogonal states can be distinguished perfectly with a single-copy measurement~\citep{Nielsen2000book}. Both of these limitations are entirely absent in $\mathcal{PT}$-symmetric quantum theory because the following two assumptions are built in: 
\begin{enumerate}
\item There exists an local quantum system described by a $\mathcal{PT}$-symmetric Hamiltonian and it can coexist with a conventional quantum system.
\item Post-measurement probability distributions are computed using conventionally normalized quantum states.
\end{enumerate}

These two assumptions are implicitly made in~\citep{Bender2007, Bender2013} and present a clear departure from standard quantum mechanics, but so far have not been tested. The existing experimental realizations of $\mathcal{PT}$-symmetric evolutions are either classical simulations or conditioned evolution in conventional quantum theory~\citep{Ruter2010, Zheng2013}. Some theoretical scrutiny has shown that a \emph{globally} $\mathcal{PT}$-symmetric system is conventional quantum mechanics in disguise with a different inner product definition, and in finite-dimensional systems $\mathcal{PT}$-symmetric Hamiltonians are actually a specific class of pseudo-Hermitian Hamiltonians in one-to-one correspondence with Hermitian Hamiltonians via a similarity transformation~\citep{Mostafazadeh2001, *Mostafazadeh2002, *Mostafazadeh2002a, *Mostafazadeh2003, *Mostafazadeh2004, *Mostafazadeh2007}. This equivalence indicates that if $\mathcal{PT}$-symmetric quantum symmetry can only describe physical systems globally then it would be unnecessary for us to consider this theory except for potentially simplifying calculations. From this point of view, whether $\mathcal{PT}$-symmetric quantum theory is a valid \emph{local} theory, i.e.\ certain subsystems are $\mathcal{PT}$ symmetric while others in general aren't~\citep{Gunther2008,Gunther2008a}, becomes a significant question.

Although $\mathcal{PT}$-symmetric systems satisfy the requirements of real energy spectrum bounded from below and probability conservation, they still must satisfy other physical limitations. Here we examine the assumptions 1.\ and 2.\ using the no-signaling conditions from special relativity: $\forall b,B,A_{\pm}$,
\begin{align}
\label{eqn:nosignal}
	\sum_a P(a,b|A_+,B)=\sum_a P(a,b|A_-,B)=P(b|B),
\end{align}
where $a$, $b$ are measurement outcomes of two space-like separated parties Alice and Bob, and $A_{\pm}$ and $B$ are different local measurements done by Alice and Bob on their respective sides. The meaning of Eq.~(\ref{eqn:nosignal}) is that Bob's local measurement-outcome probability distribution is unaffected by Alice's choice of local measurements. 

The main result of this paper is that any locally $\mathcal{PT}$-symmetric system will in general violate Eq.~\ref{eqn:nosignal} if both of the assumptions 1.\ and 2.\ are true. This greatly restricts the realm of interest for this theory to a curious form of effective theory, unless the astonishing and highly unlikely possibility of superluminal communication is realized.

\emph{$\mathcal{PT}$-symmetric Hamiltonians} ---
A Hamiltonian $H$ is $\mathcal{PT}$ symmetric if it commutes with the parity $\mathcal{P}$ and time reversal $\mathcal{T}$ operators. In a two-level system, $\mathcal{P}$ is defined by the Pauli $\sigma_x$ matrix and $\mathcal{T}$ is defined by complex conjugation; a non-trivial example of a $\mathcal{PT}$-symmetric Hamiltonian is
\begin{align}
\label{eqn:hamiltonian}
	H=&s
	\begin{pmatrix}
		i\sin\alpha &1\\
		1 &-i\sin\alpha
	\end{pmatrix},\quad s,\,\alpha\in\mathbb{R}\,,
\end{align}
where $s$ is a scaling constant and $\alpha$ is called the non-Hermiticity of $H$~\citep{Gunther2008}. When $\alpha=0$, $H$ is a Hermitian Hamiltonian. The (right) eigenvalues, $E_\pm=\pm s\cos\alpha$, are real when $|\alpha|<\pi/2$, corresponding to the (right) eigenstates
\begin{align}
	|E_+(\alpha)\rangle&=\frac{e^{i\alpha/2}}{\sqrt{2\cos\alpha}}
		\begin{pmatrix}
			1\\
			e^{-i\alpha}
		\end{pmatrix},\notag\\
	|E_-(\alpha)\rangle&=\frac{ie^{-i\alpha/2}}{\sqrt{2\cos\alpha}}
		\begin{pmatrix}
			1\\
			-e^{i\alpha}
		\end{pmatrix}.\notag
\end{align}
These states are not orthogonal to each other in conventional quantum theory. When $\alpha=\pm\pi/2$, they become the same state, and this is the $\mathcal{PT}$ symmetry-breaking point.

The time-evolution operator for such a system is, following~\citep{Bender2007}, given by
\begin{align*}
	U(t)\equiv e^{-itH}=\frac{1}{\cos\alpha}
	\begin{pmatrix}
		\cos(t'-\alpha) &-i\sin t'\\
		-i\sin t'        &\cos(t'+\alpha)
	\end{pmatrix},
\end{align*}
where $t'\equiv \frac{\Delta E}{2} t$, $\Delta E = E_+-E_-$, and $\hbar=1$. 

\emph{Violation of no-signaling condition} ---
Suppose that two space-like separated parties, Alice and Bob, want to transmit information without using any classical communication. They are permitted to discuss their communication protocol and share a maximally entangled state $|\psi\rangle=\frac{1}{\sqrt{2}}(|+_x+_x\rangle+|-_x-_x\rangle)$ beforehand, where $|\pm_k\rangle$ are eigenstates of the Pauli matrices $\sigma_k$, $k\in \{x,y,z\}$. If Alice has a local $\mathcal{PT}$-symmetric quantum system $H$ and it does not interact with any subsystem on Bob's side, then the total Hamiltonian describing the composite system is $H_{\mathrm{tot}}=H\otimes I$, where $I$ is the identity operator. This prescription also holds in $\mathcal{PT}$-symmetric systems because the identity operator keeps the same form in both kinds of quantum theory. According to the process of the gedanken experiment in~\citep{Bender2007} and the previous two assumptions, if Alice first uses the operator $A_+ = I$ or $A_- = \sigma_x$ with respect to the information she wants to send and sets the time of evolution to $\tau=\pi/\Delta E$, the joint final states are 
\begin{align*}
|\psi_f^\pm\rangle&=[U(\tau) A_\pm \otimes e^{-iIt}I]|\psi\rangle\notag\\
	&\propto\frac{1}{\sqrt{2}}\left[e^{i\phi_+}
	\begin{pmatrix}
	1\\
	ie^{-i\epsilon}
	\end{pmatrix}|+_x\rangle 
	\pm e^{i\phi_-}
	\begin{pmatrix}
	1\\
	ie^{i\epsilon}
	\end{pmatrix}|-_x\rangle\right]\,,
\end{align*}
where $e^{i\phi_{\pm}}=\frac{\sin\alpha\mp i}{\sqrt{1+\sin^2\alpha}}$ and $e^{i\epsilon}=\frac{-2\sin\alpha+i\cos^2\alpha}{1+\sin^2\alpha}$. Here we note that the normalization constants have been renormalized in the way of conventional quantum mechanics, since in the end Bob will measure it using conventional quantum mechanics. In the extreme case that $\alpha\rightarrow -\pi/2$, the respective states that Bob holds are
\begin{align*}
	\rho_B^\pm\equiv\mbox{Tr}_A(|\psi_f^\pm\rangle\langle\psi_f^\pm|)=|\pm_y\rangle\langle\pm_y|,
\end{align*}
Thus from the measurement outcomes Bob can learn the information Alice wants to transmit. 

In fact, this result continues to hold for all $\alpha$ that yield a non-Hermitian $H$. Following the previous protocol, Alice and Bob both measure their systems with the conventional quantum projectors $|\pm_y\rangle\langle\pm_y|$, which gives the joint probabilities 
\begin{align*}
	P(a,b|A_{\pm},B)=\langle\psi_f^\pm|(|a\rangle\langle a|\otimes|b\rangle\langle b|)|\psi_f^\pm\rangle,
\end{align*} 
where the possible outcomes of $a$ and $b$ are $+_y$ or $-_y$. After a simple calculation, we have the two marginal probabilities
\begin{align*}
	\sum_{a=\pm_y}P(a,+_y| A_\pm ,B)=\frac{1}{2}[1\pm\cos\epsilon\sin(2\phi_+-\epsilon)]\,.
\end{align*}
The two equations are the same only when $\cos\epsilon=0$, which implies that the no-signaling condition is always violated unless $\alpha=2\pi n$, i.e.\ the system used by Alice is Hermitian.

%-----------------------------------------------------------------------------------------------------------------------%
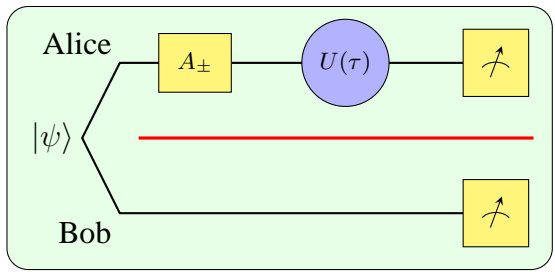
\begin{figure}[th!]
\centering
% Set project root file
% !TEX root = PTsymmetric.tex
\tikzsetnextfilename{process}
\begin{tikzpicture}[PT/.style={inner sep=1.5mm,circle,draw,fill=blue!30},
			QM/.style={inner sep=2.5mm,rectangle,draw,fill=yellow!65},
			bg/.style={inner sep=2mm,circle,draw,fill=green!10}]
%	\draw[color=gray,scale=.5] (0,0) grid (14,6);
	\draw[color=red,very thick] (1.75,1.5) -- ++(5.25,0);
	\draw[color=black, thick] (6.5,.5) -- ++(-5,0) -- ++(-.5,1) -- ++(.5,1) -- ++(5,0);
	\draw (1.5,.5) node[anchor=north east]{{\large Bob}}
		 (1.5,2.5) node[anchor=south east]{{\large Alice}};
	\draw (2.5,2.5) node[QM]{$A_\pm$} ++(2,0) node[PT]{$U(\tau)$};
	\draw (1,1.5) node[anchor=east]{{\large $|\psi\rangle$}};
	\node[QM,minimum size=1.35em] at (6.5,2.5) {
		\begin{tikzpicture}
			\draw[->,>=stealth,shorten >=-1pt,thin] (-.07,-.18) -> (.07,.2);
			\draw (0,.02) arc (90:160:.6em and .35em);
			\draw (0,.02) arc (90:20:.6em and .35em);
		\end{tikzpicture}
	};
	\node[QM,minimum size=1.35em] at (6.5,.5) {
		\begin{tikzpicture}
			\draw[->,>=stealth,shorten >=-1pt,thin] (-.07,-.18) -> (.07,.2);
			\draw (0,.02) arc (90:160:.6em and .35em);
			\draw (0,.02) arc (90:20:.6em and .35em);
		\end{tikzpicture}
	};
	\begin{pgfonlayer}{background} 
		\filldraw[bg,rounded corners=8pt] (0,-.25) rectangle (7.25,3.25) {};
	\end{pgfonlayer}
\end{tikzpicture}
\caption{\label{fig:process} Alice and Bob initially share a maximally entangled state $|\psi\rangle$ and are spacelike separated (red line). The circled operators are $\mathcal{PT}$ symmetric, while rectangular ones are conventional operators; the identity gate (a wire) is the same for both theories. Alice's initial choice of $A_\pm$ is followed by $\mathcal{PT}$-symmetric time evolution $U(\tau)$. A projective measurement at the end leads to superluminal signaling.}
\end{figure}
%-----------------------------------------------------------------------------------------------------------------------%

\textit{Discussion} ---
We have demonstrated that the two assumptions made for accomplishing ultrafast quantum processes and discrimination of non-orthogonal states will lead to the violation of the no-signaling condition. This violation happens not only for the Hamiltonian $H$ in Eq. (\ref{eqn:hamiltonian}), but for all $2\times2$ (nontrivial) $\mathcal{PT}$-symmetric Hamiltonians with even time-reversal symmetry $\mathcal{T}^2=+1$, which follows by a suitable unitary transformation on $H$. By a simple embedding argument, any nontrivial $\mathcal{PT}$-symmetric $H$ of higher dimensions will also violate no signaling, so the result is quite general.

Our result seems to lead $\mathcal{PT}$-symmetric quantum theory into the following trichotomy of possible situations:

\textbf{a.} The first assumption is incorrect. $\mathcal{PT}$-symmetric Hamiltonians are not local and the model of $\mathcal{PT}$-symmetric quantum theory does not completely describe a real physical system, or it cannot be regarded as a real physical system.

\textbf{b.} The first assumption is true but the second assumption is incorrect. Thus, the rules describing how the standard and $\mathcal{PT}$-symmetric theories transition between each other must be modified to avoid superluminal signaling. To our knowledge, two ways are known to establish a one-to-one transformation between the states in the standard frame and in the $\mathcal{PT}$ frame. The first one is our second assumption, and the second one is the similarity transformation discovered by Mostafazadeh~\citep{Mostafazadeh2001, *Mostafazadeh2002, *Mostafazadeh2002a, *Mostafazadeh2003, *Mostafazadeh2004, *Mostafazadeh2007}. However, the first way, as we have  already shown, will violate no-signaling, and the second way makes ultrafast time evolution and discrimination of non-orthogonal quantum states impossible, essentially reducing it to standard quantum mechanics. Furthermore, this transition theory should include the interactions between two separated parties, otherwise the violation of the no-signaling condition cannot be explained since all the operations in Fig.~(\ref{fig:process}) are local.

\textbf{c.} Both assumptions are true, and $\mathcal{PT}$-symmetric systems give us the ability to do all of these powerful applications, including superluminal signaling. However, this situation seems to be by far the most unlikely one.

The central problem with local $\mathcal{PT}$-symmetric theories (generously assuming scenario \textbf{b}.)\ is that the renormalization caused by the transition between two different systems is a \emph{nonlinear} map which would cause superluminal signaling~\citep{Cavalcanti2012}, and in general other highly implausible scenarios such as solving \#P problems in polynomial time~\cite{Abrams1998}. In fact, our conclusions persist if we normalize in the $\mathcal{PT}$-symmetric inner product instead, since the nonlinearity comes from the \emph{relative} distortion of the state space between the two theories. Nonlinear quantum theory has been debated for a long time~\citep{Weinberg1989, *Weinberg1989a, *Gisin1990, *Gisin1989}, and the possibility of nonlinear time evolution is not completely ruled out. Although sometimes these symptoms can be ameliorated~\cite{Kent2005}, it seems that the medicine of additional assumptions is worse than than the original ailment. 

Finally, while in our view these results essentially kill any hope of $\mathcal{PT}$-symmetric quantum theory as a fundamental theory of nature, it could still be useful as an effective model or as a purely mathematical problem-solving device. 

\acknowledgments
We thank Eric Cavalcanti, Eric Chitambar, and Nick Menicucci for useful discussions. STF was supported by the ARC Centre of Excellence for Engineered Quantum Systems CE110001013. MH was supported by the UTS ChancellorÕs postdoctoral research fellowship and UTS Early Career Researcher Grants Scheme. YCL appreciates the hospitality of CQCIS during his stay in Sydney.

\bibliography{PTsymmetric}

\end{document}